\documentclass[reprint,amsmath,amssymb,aps]{revtex4-1}

\renewcommand{\vec}[1]{\mathbf{#1}}
\usepackage{graphicx}
\usepackage{braket}
\usepackage{wasysym}
\usepackage{color}
\usepackage{nicefrac}
\usepackage{sidecap}
\usepackage{subfigure}
\usepackage[pdftex]{hyperref}
\usepackage[normalem]{ulem}

\renewcommand{\eqref}[1]{Eq.~(\ref{eq:#1})}

\newcommand{\citesup}[1]{\textsuperscript{\citealp{#1}}}
\newcommand{\textsubscript}[1]{$_{\text{#1}}$}

\definecolor{darkviolet}{rgb}{0.58, 0.0, 0.83}
\def\a{s}
\def\b{s}
\newcommand{\add}[1]{\if\a\b{{\color{red} #1}}\else{#1}\fi} 
\newcommand{\del}[1]{\if\a\b{{\color{blue} \sout{#1}}}\else{}\fi}
\newcommand{\com}[1]{\if\a\b{{\color{darkviolet} #1}}\else{#1}\fi}
 
\begin{document}

\title{Dynamical laser spike processing}

\author{Bhavin J. Shastri}\thanks{These authors contributed equally to this work.}
\author{Mitchell A. Nahmias}\thanks{These authors contributed equally to this work.}
\author{Alexander N. Tait}\thanks{These authors contributed equally to this work.}
\author{Alejandro W. Rodriguez}
\author{Ben Wu}
\author{Paul R. Prucnal}
\email{prucnal@princeton.edu}
\affiliation{Department of Electrical Engineering, Princeton University, Princeton, New Jersey 08544, USA}
\date{\today}

\begin{abstract}
Novel materials and devices in photonics have the potential to revolutionize optical information processing, beyond conventional binary-logic approaches. Laser systems offer a rich repertoire of useful dynamical behaviors, including the excitable dynamics also found in the time-resolved ``spiking'' of neurons. Spiking reconciles the expressiveness and efficiency of analog processing with the robustness and scalability of digital processing. We demonstrate that graphene-coupled laser systems offer a unified low-level spike optical processing paradigm that goes well beyond previously studied laser dynamics. We show that this platform can simultaneously exhibit logic-level restoration, cascadability and input-output isolation---fundamental challenges in optical information processing. We also implement low-level spike-processing tasks that are critical for higher level processing: temporal pattern detection and stable recurrent memory. We study these properties in the context of a fiber laser system, but the addition of graphene leads to a number of advantages which stem from its unique properties, including high absorption and fast carrier relaxation. These could lead to significant speed and efficiency improvements in unconventional laser processing devices, and ongoing research on graphene microfabrication promises compatibility with integrated laser platforms.

\end{abstract}

\maketitle


Recently, there has been a pertinacious exploration of the unifying boundaries between information communication (dominated by optics) and information processing (dominated by electronics) in the same medium. In the context of information processing, nonlinear dynamical systems\citesup{strogatz2014nonlinear,Appeltant:2011aa,Jaeger02042004} have been receiving considerable attention due to their isomorphism to biological networks. Compared to binary-logic based methods implemented on standard von Neumann architectures, unconventional processing paradigms that are neuroinspired\citesup{Merolla08082014,Jaeger02042004,Hasler:2013,Indiveri:2011} are relatively more effective for solving certain tasks, such as pattern analysis, decision-making, optimization, and learning. A sparse coding scheme, called spiking,\citesup{Izhikevich:2003,Ostojic:2014aa} has recently been recognized by the neuroscience community as an important neural coding strategy for information processing.\citesup{Ostojic:2014aa,Kumar:2010aa,Diesmann:1999aa,Borst:1999aa} The continued evolution of photonic technologies has reawakened interest for a relentless search in neuro-inspired optical information processing\citesup{Vandoorne:2014aa,Brunner:2013aa,Appeltant:2011aa,Woods:2012aa,Sorrentino:15} to complement and enable new opportunities\citesup{Caulfield:2010aa,Tucker:2010aa} and potentially bridge the gap with information communication in the same substrate.\citesup{Miller:2010aa} 

In this manuscript, we provide the first unified, experimental demonstration of low-level spike processing\citesup{Izhikevich:2003,Ostojic:2014aa} functions in an optical platform. We exploit unconventional (excitable) dynamical properties of graphene laser systems to demonstrate the following features (in a fiber-based prototype) which are key impediments to optical computing\citesup{Caulfield:2010aa,Tucker:2010aa,Miller:2010aa}: logic-level restoration, cascadability, and input-output isolation. Although a number of approaches have demonstrated these properties separately,\citesup{Coomans:2011,Hurtado:2012,Barbay:2011} no reported devices have simultaneously demonstrated these critical functionalities together in a single device.\citesup{Caulfield:2010aa,Tucker:2010aa,Miller:2010aa}
Our experimental prototype also possesses properties useful for processing tasks, including temporal integration, and sharp thresholding, leading to a very simple temporal classifier.\citesup{Kravtsov:2011} We include a simulation model that explains all of the observed behaviors: integration, thresholding, refractoriness, and pulse generation. We also propose and simulate an analogous integrated device structure that exhibits the same dynamics in $\lesssim\mathrm{mm^2}$ footprints. Scaling down the cavity length and overall size (by factors of millions) allows an integrated graphene excitable laser to exhibit dynamics on the order of ps timescales. Our model draws inspiration from novel insights in event-based information representation, dynamical excitability, and the unique material properties of graphene.
\begin{figure*}[t]
\centering\includegraphics[width=1.5\columnwidth]{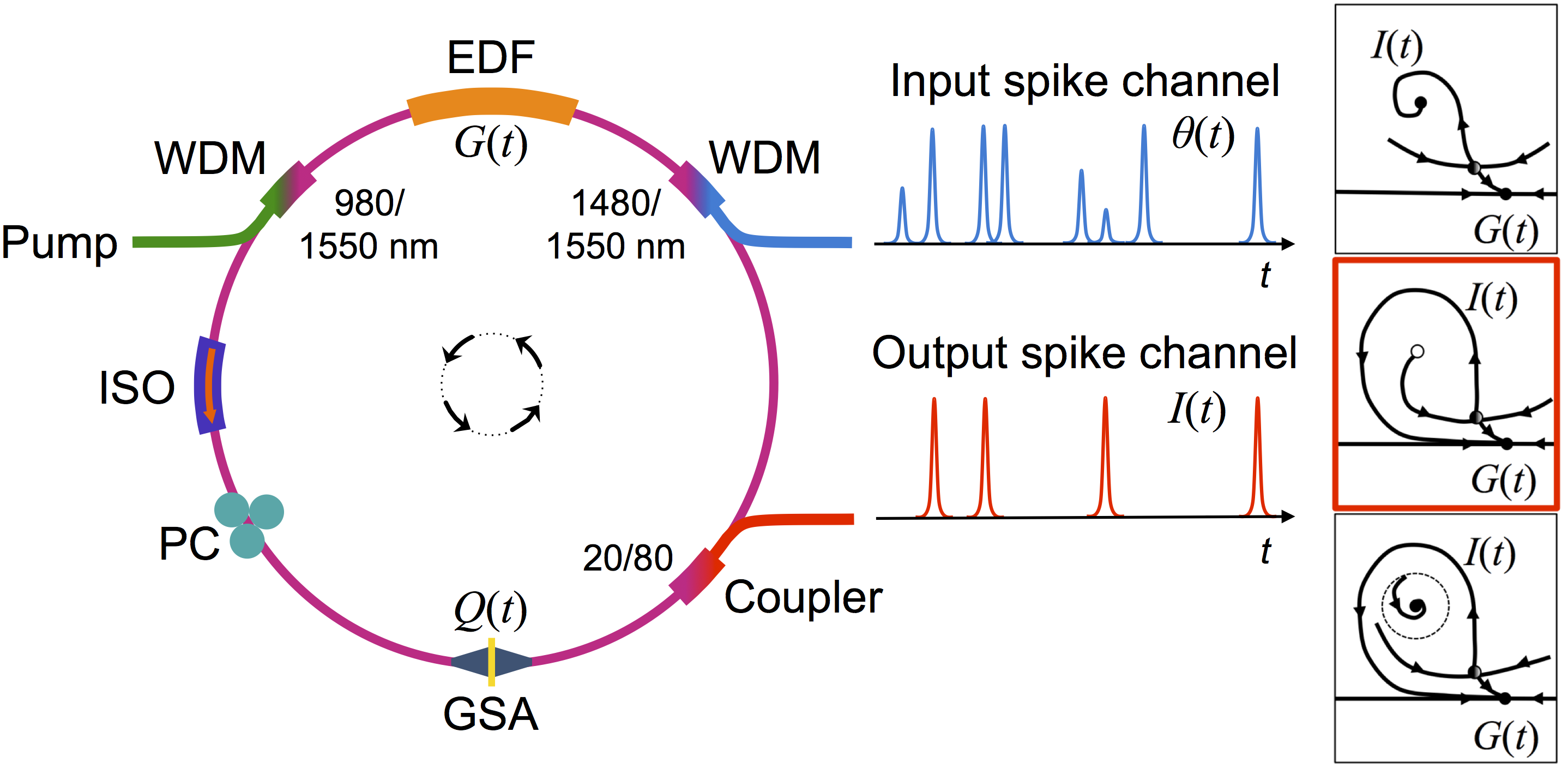}
\caption{\label{fig:fiber_laser} \textbf{Graphene excitable fiber laser.} The cavity consists of a chemically synthesized (see Methods) graphene SA (GSA) sandwiched between two fiber connectors with a fiber adapter and a 75-cm long highly doped erbium-doped fiber (EDF) as the gain medium. The EDF is pumped with a 980 nm laser diode via a 980/1550 nm wavelength-division multiplexer (WDM). An isolator (ISO) ensures unidirectional propagation. A polarization controller (PC) maintains a given polarization state, improving output pulse stability.  The 20\% port of an optical coupler provides the laser output at 1560 nm. To induce perturbations to the gain, 1480 nm excitatory pulses are incident on the system via a 1480/1550 nm WDM. These analog inputs---from other excitable lasers, for example---are directly modulated with an arbitrary waveform generator. Right: illustration of different possible phase-space dynamics associated with the laser intensity $I$ and gain $G$ of the system (see Text), as the various physical parameters (pump power, length of cavity, absorption) are varied. The desired excitable behavior, corresponding to the second phase-space schematic (red), is achieved when the parameter regimes drive the system toward a so-called homoclinic bifurcation.\citesup{strogatz2014nonlinear}}
\end{figure*}

Spiking is a sparse coding scheme with firm code-theoretic justifications.\citesup{Sarpeshkar:1998,Thorpe:2001,Maass:2002} Information is encoded in the temporal and spatial relationships between short pulses (or `spikes'). Spike codes---which are digital in amplitude but analog in time---exhibit the expressiveness and efficiency of analog processing with the robustness of digital communication. 
Spikes are typically received and generated by nonlinear dynamical systems, and can be represented and processed dynamically through excitability---a far-from-equilibrium nonlinear dynamical mechanism underlying all-or-none responses to small perturbations.\citesup{Hodgkin:1952} Excitable systems possess unique regenerative properties and have been employed for sensing microparticles with an optical torque wrench\citesup{Pedaci:2011aa} and image processing utilizing a photosensitive Belousov--Zhabotinsky reaction.\citesup{Kuhnert:1989aa} In the context of spike processing, excitable laser systems\citesup{Turconi:2013,Hurtado:2012,Barbay:2011} have been studied with the tools of bifurcation theory.\citesup{VanVaerenbergh:12,Coomans:2011,Romeira:2013}. Many dynamical systems that are explored are closely tied to underlying device physics, and, as such, the search for useful systems of this kind often involves novel materials. 

Our approach exploits the unique properties of graphene, whose remarkable electrical and optical properties have enabled several disruptive technologies.\citesup{Novoselov:2012aa,Bonaccorso:2010aa,Bao:2012} 
Graphene transistors are poised to be smaller and faster compared to their silicon counterparts,\citesup{Schwierz:2010aa,Lin05022010} but poor on/off current ratios resulting from a zero bandgap poses a serious challenge for conventional digital logic. Instead of using graphene's electrical properties as an active element in conventional processing applications, we exploit its passive and unique optical properties to enable unconventional processing. 
Since its emergence as a new type of saturable absorber (SA), graphene has been rigorously studied in the context of passive mode locking and Q-switching,\citesup{Martinez:2013aa,Bao:2009,Sun:2010,Xing:10} and has been preferred over the widely used semiconductor saturable absorbers\citesup{Keller:1996} due to its high saturable absorption to volume ratio.\citesup{Novoselov:2012aa} Graphene possesses a number of other important advantages that are particularly useful in the context of processing, including a very fast response time, wideband frequency tunability (useful for wavelength division multiplexed networks), and a tunable modulation depth. Furthermore, graphene also has a high thermal conductivity and damage threshold compared to semiconductor absorbers.

This work experimentally validates the theoretically discovered\citesup{Nahmias:2013,Nahmias:2014} dynamical isomorphism between semiconductor photocarriers and neuron biophysics, along with recent predictions of spike processing enabled by graphene.\citesup{Shastri:oqe:2014,Shastri:nusod:2013} Ongoing research on graphene microfabrication\citesup{Novoselov:2012aa,Bonaccorso:2010aa,Bao:2012,Grigorenko:2012aa} may make it a standard technology accessible in integrated laser platforms, which, together with a suitable networking platform,\citesup{Tait:jlt:2014} could lead to a scalable platform for optical computing.\citesup{Caulfield:2010aa,Tucker:2010aa,Miller:2010aa}%

\begin{figure*}[t]
\centering
\centering\includegraphics[width=1.5\columnwidth]{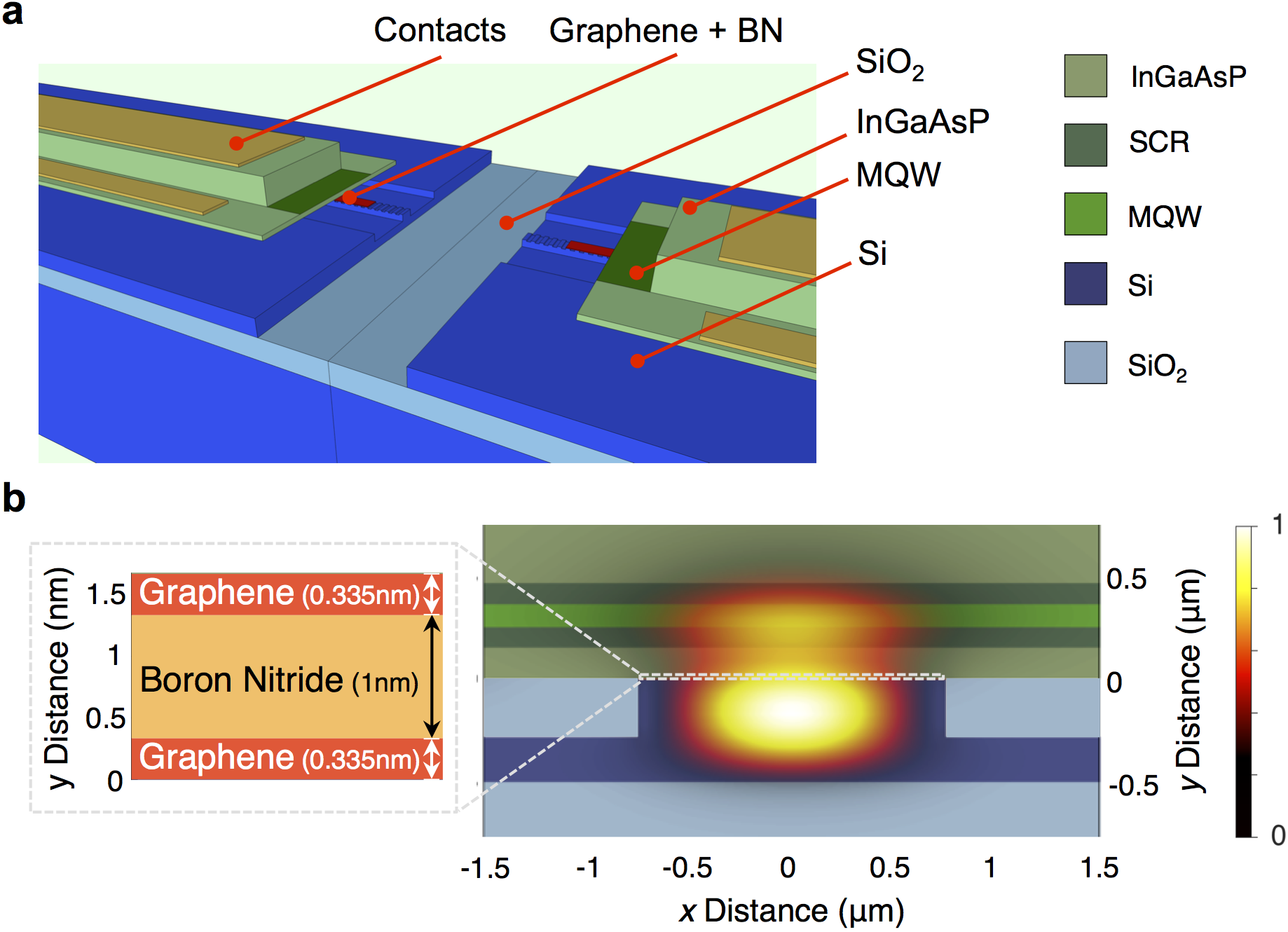}
\caption{\label{fig:integrated_laser} \textbf{Proposed integrated graphene excitable laser.} \textbf{(a)} Architecture of the hybrid InGaAsP-graphene-silicon evanescent laser (not to scale) with a terraced cutaway of the center. The device comprises a III-V epitaxial structure with multiple quantum well (MQW) region bonded to a low-loss silicon rib waveguide that rests on a silicon-on-insulator (SOI) substrate with a sandwiched heterostructure of two monolayer graphene sheets and an hexagonal boron nitride (hBN) spacer. Note that, unlike the fiber laser of Fig.~\ref{fig:fiber_laser}, the gain section of this structure is  electrically pumped. The cavity and waveguide are formed by the presence of a half-wavelength grating in the silicon. \textbf{(b)} Cross-sectional profile of the excitable laser with an overlaid electric field (E-field) intensity $|\vec{E}|^2$ profile. The optical mode of the laser lies predominately in the silicon waveguide with a small portion of the mode overlapping the QWs of the III-V structure for optical gain and the 2D materials heterostructure for absorption. The silicon waveguide has a width, height, and rib etch depth of 1.5$\mu$m, 500nm, and 300nm, respectively. The calculated overlap of the optical mode with the silicon waveguides is 0.558 while there is a 0.00046 overlap in the graphene sheets and 0.043 overlap in the QWs. The E-field intensities are calculated at a wavelength of 1.5$\mu$m. Graphene's thickness of 0.335nm and absorption coefficient of 301,655 cm$^{-1}$ are used for the simulations (see Methods).
}
\end{figure*}

\section*{Results}\label{sec:results}
{\flushleft{\bf Dynamical model.}} The dynamical system underlying the behavior of our spike processing unit is a gain-absorber cavity model, describing single mode lasers with gain and SA sections. Despite its simplicity, it can exhibit a large range of dynamical behaviors,\citesup{Dubbeldam:1999} and has been investigated in various contexts as the basis for an optical processor.\citesup{Selmi:2014} The system, in its simplest form, can be described using the following undimensionalized equations\citesup{Barbay:2011, Nahmias:2013}:
\begin{subequations}
\label{eq:yamada}
\begin{align}
\label{eq:yamadaG} \dot G(t) &= \gamma_G [ A - G(t) - G(t) I(t)] + \theta(t) \\
\label{eq:yamadaQ} \dot Q(t) &= \gamma_Q [B - Q(t) - a Q(t) I(t)] \\
\label{eq:yamadaI} \dot I(t) &= \gamma_I [G(t) - Q(t) - 1] I(t) + \epsilon f(G)
\end{align}
\end{subequations}
where $G(t)$ models the gain, $Q(t)$ the absorption, and $I(t)$ the laser intensity. $A$ is the gain bias current, $B$ is the absorption level, $\gamma_G$ is the gain relaxation rate, $\gamma_Q$ is the absorber relaxation rate, $\gamma_I$ is the inverse photon lifetime, and $a$ is a differential absorption relative to the gain factor. We represent the spontaneous noise contribution to intensity via $\epsilon f(G)$, for small $\epsilon$, and time-dependent input perturbations as $\theta(t)$.

When the dynamics of pulse generation are fast compared to the dynamics of the gain medium, one can compress the internal dynamics and obtain an instantaneous pulse-generation model\citesup{Nahmias:2013}:
\begin{subequations}
\begin{align}
	\frac{dG(t)}{dt}&=-\gamma_G(G(t)-A)+\theta(t);\\
	&\text{if $G(t)>G_{\mathrm{thresh}}$ then}\\\nonumber
	&\text{release a pulse, and set $G(t)\rightarrow G_{\mathrm{reset}}$}
\end{align}
\end{subequations}
where the input $\theta(t)$ can include spike inputs of the form $\theta(t) = \sum_i \delta_i (t - \tau_i)$ for spike firing times $\tau_i$, $G_{\mathrm{thresh}}$ is the gain threshold, and $G_{\mathrm{reset}}\sim 0$ is the gain at transparency.

This system is analogous to a leaky integrate-and-fire (LIF) neuron model, commonly employed in computational neuroscience for modeling biological neural networks. Although it is one of the simpler spike-based models, the LIF model is capable of universal computations,\citesup{Maass:1997} and the transmission of information through spike timings.\citesup{Strong:1998} The gain-absorber system has been predicted to exhibit cascadability, logic-level restoration, and input-output isolation,\citesup{Nahmias:2013} satisfying the basic criteria for optical computing.\citesup{Miller:2010aa}

\begin{figure*}[t]
\centering\includegraphics[width=2\columnwidth]{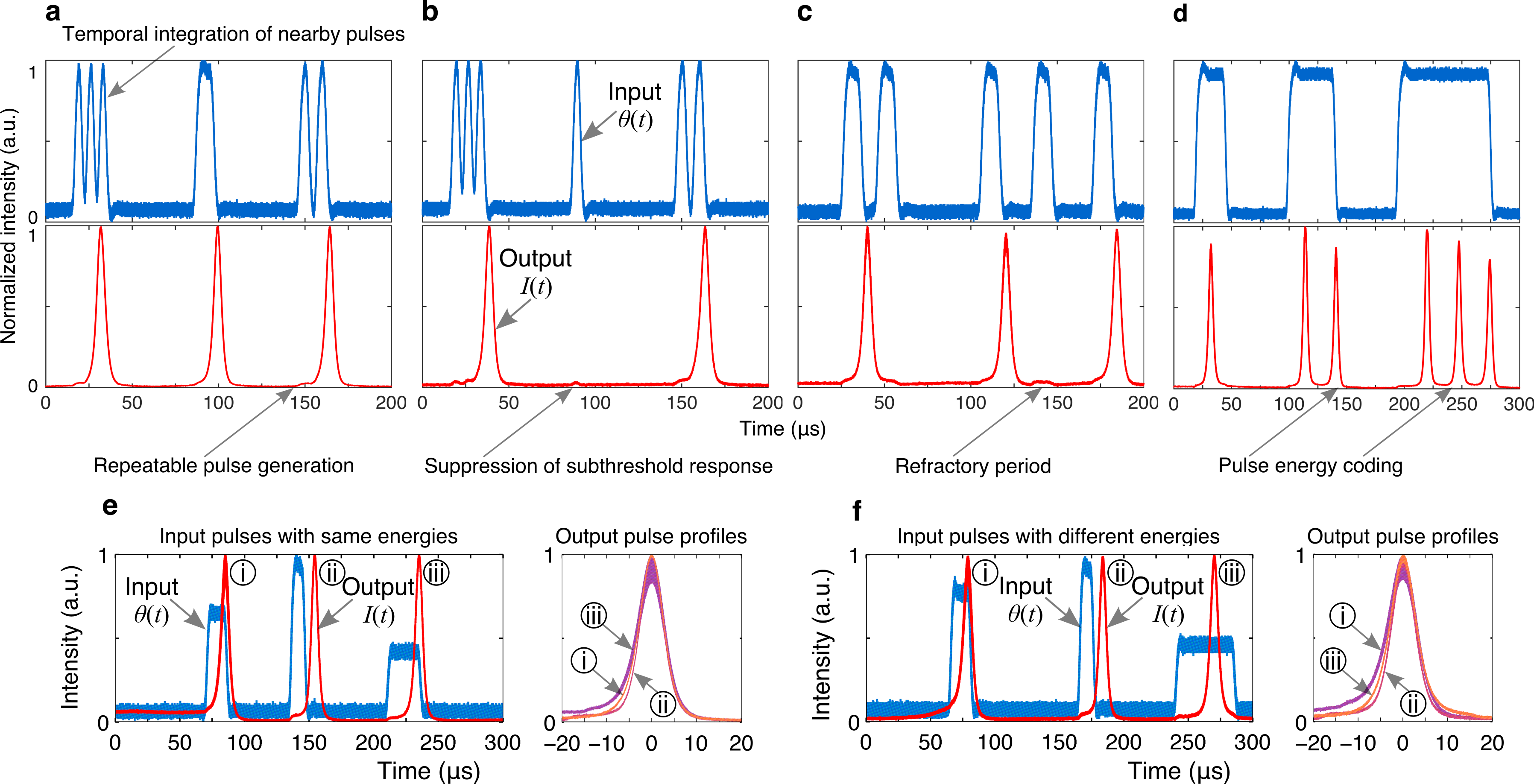}
\caption{\label{fig:excitable_dynamics} \textbf{Excitable dynamics of the graphene fiber laser.} Note the blue and red curves correspond to input and output pulses, respectively. \textbf{(a-c)} Excitatory activity (temporal integration of nearby pulses) can push the gain above the threshold (measured to be $\approx 275$ nJ), releasing spikes. Depending on the input signal, the system can lead to (a) repeatable pulse generation, or suppressed response due to the presence of either (b) sub-threshold input energies (integrated power $\int |\theta(t)|^2 dt$) or (c) a refractory period during which the gain recovers to its resting value and the laser is unable to pulse (regardless of excitation strength). \textbf{(d)} Typical bursting behavior i.e. emission of doublets (two spikes) and triplets (three spikes) when a strong input drives the system over the threshold to fire repetitively. \textbf{(e-f)} Restorative properties (repeatable pulse reshaping) of spike processing; inputs with either the (e) same or (f) different energies. Test conditions: (a) three 4~$\mu$s input pulses separated by 3~$\mu$s followed by a 10~$\mu$s pulse after 52~$\mu$s delay, and two 5.5~$\mu$s pulses separated by 4.5~$\mu$s after 50~$\mu$s delay; (b) three 4~$\mu$s input pulses separated by 3~$\mu$s followed by a 4.5~$\mu$s pulse after 52~$\mu$s delay, and two 5.5~$\mu$s pulses separated by 4.5~$\mu$s after 50~$\mu$s delay; (c) five 10~$\mu$s input pulses separated by 10, 50, 20, and 25~$\mu$s delays, respectively; (d) three input pulses with widths 25, 40, and 80~$\mu$s separated by 55~$\mu$s; in (a-d) the 980~nm pump is biased at 60.4~mW. (e) three input pulses with energies 252, 264, and 256~nJ separated by 85 and 65~$\mu$s, result in outputs with pulses widths (i) 7.4, (ii) 6.1, and (iii) 6.9~$\mu$s, respectively; 980~nm pump is biased at 61~mW. (f) three input pulses with energies of 265 nJ separated by 50 and 65~$\mu$s, result in outputs with pulse widths (i) 6.3, (ii) 6, and (iii) 6.5~$\mu$s; 980~nm pump is biased at 67.6~mA.}
\end{figure*}

{\flushleft{\bf Excitable Laser Systems.}} Our demonstration of spike processing is based on a graphene fiber ring laser platform (Fig.~\ref{fig:fiber_laser}). For comparison, we also perform numerical simulations of an analogous proposed integrated device (Fig.~\ref{fig:integrated_laser}). Both devices along with their respective simulation models and parameters are described in detail in the Methods section. The fiber ring laser contains an erbium doped fiber amplifier (gain section) and liquid exfoliated graphene (absorber section), interacting with one another in a fiber ring (cavity). The ring laser pulses periodically if driven above a threshold, modulated by the passive saturation of graphene absorption. This behavior has been studied in the context of high power, wideband passively Q-switched lasers, for which graphene has many favorable properties.\citesup{Popa:2011}

The integrated device contains electrically pumped quantum wells (gain section), two sheets of graphene (absorber section), and a distributed feedback-grating (section). In this design, we consider a hybrid silicon III-V laser platform in which the graphene layers are sandwiched in between the silicon and III-V layers. The hybrid III-V platform is highly scalable and amenable to both passive and active photonic integration\citesup{Fang:2007}. The integrated device is capable of exhibiting the same behaviors as the fiber prototype, but on a much faster time scale and with lower pulse energies. Figure S\ref{fig:q_switching} compares the pulse repetition rate and pulse widths as a function of input power between the integrated device and fiber laser. In both cases, the rate of output pulses depends monotonically on the amount of power being consumed. This has many behavioral similarities with the behavior of rate neurons, which code information through spike frequency modulation.\citesup{fries2001modulation} Although both lasers consume similar amounts of power, the integrated device pulses $\sim10^6$ times faster. This corresponds to a $\sim10^6$ times decrease in the energy consumed per pulse. The devices (and their respective simulation models) are described in more detail in the Methods section.

{\flushleft{\bf Excitability.}} We demonstrate that both the fiber ring laser and the integrated device are excitable and capable of performing spike processing tasks. Excitability is defined by three main criteria: (i) an unperturbed system rests at a stable equilibrium; (ii) a perturbation above the excitability threshold triggers a large excursion from this equilibrium; (iii) the system then settles back to the attractor in what is called the refractory period, after which the system can be excited again.\citesup{Krauskopf:2003}

\begin{figure*}[htbp]
\centering\includegraphics[width=1.75\columnwidth]{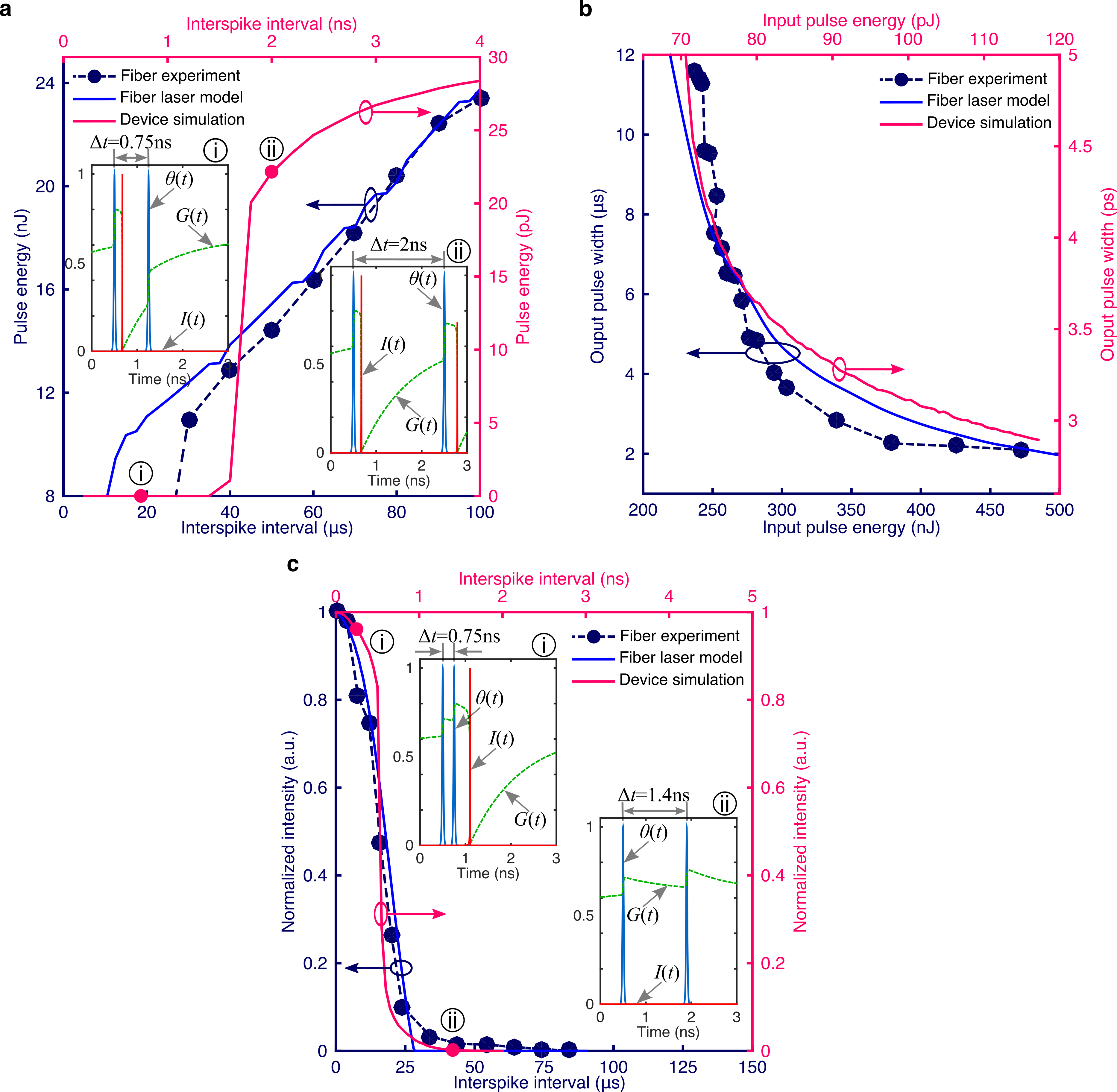}
\caption{\label{fig:excitability_properties} \textbf{Second-order properties of excitability.} \textbf{(a)} Response of the excitable laser (output pulse energy) to a second input pulse as a function of the interspike interval between two identical (first and second) excitatory pulses after the first pulse has triggered an excitable response. Both the fiber based excitable laser experimental results and the integrated excitable laser simulations exhibit absolute refractory periods (where the second pulse produces no output) and relative refractory periods (where the output response is reduced from its resting value). However, the later operates $\sim 10^3$ faster (ns compared to $\mu$s) with $\sim 10^3$ lower output pulse energies (pJ compared to nJ). Insets show the transient dynamics of the integrated excitable laser, i.e. the intensity $I(t)$ and recovery of gain carriers $G(t)$ as a result of input signal $\theta(t)$, before (i) and after (ii) its refractory period. \textbf{(b)} Excitable laser's output behavior in response to a single input pulse with different energies. The integrated excitable laser simulations also follows a similar relationship profile but with output pulse widths $\sim 10^6$ smaller (ps compared to $\mu$s). \textbf{(c)} Response of the excitable laser when implemented a coincidence detector: the excitable laser is biased such that it will not fire unless two excitatory pulses are temporally close together. Output response is strongly dependent on the temporal correlation of two inputs. Average input power is kept constant with changing pulse interval. Insets show the simulated pulse dynamics for the integrated laser for pulses that are (i) closer vs (ii) farther apart.} 
\end{figure*}

Figures~\ref{fig:excitable_dynamics}a--c demonstrates excitability within the fiber ring laser. In this system, an excitatory pulse increases the carrier concentration within the gain region by an amount proportional to its energy (integrated power) through gain enhancement. Beyond some threshold excitation energy, the absorber is saturated, resulting in the release of a pulse. This is followed by a relative refractory period during which the arrival of a second excitatory pulse is unable to cause the laser to fire as the gain recovers. The system is also capable of emitting spike doublets or triplets (see Fig.~\ref{fig:excitable_dynamics}d) in which the inter-spike timing encodes information about the pulse width and amplitude, a useful encoding scheme for selective activation.\citesup{Izhikevich:2003b} 

\begin{figure*}[t]
\centering\includegraphics[width=1.5\columnwidth]{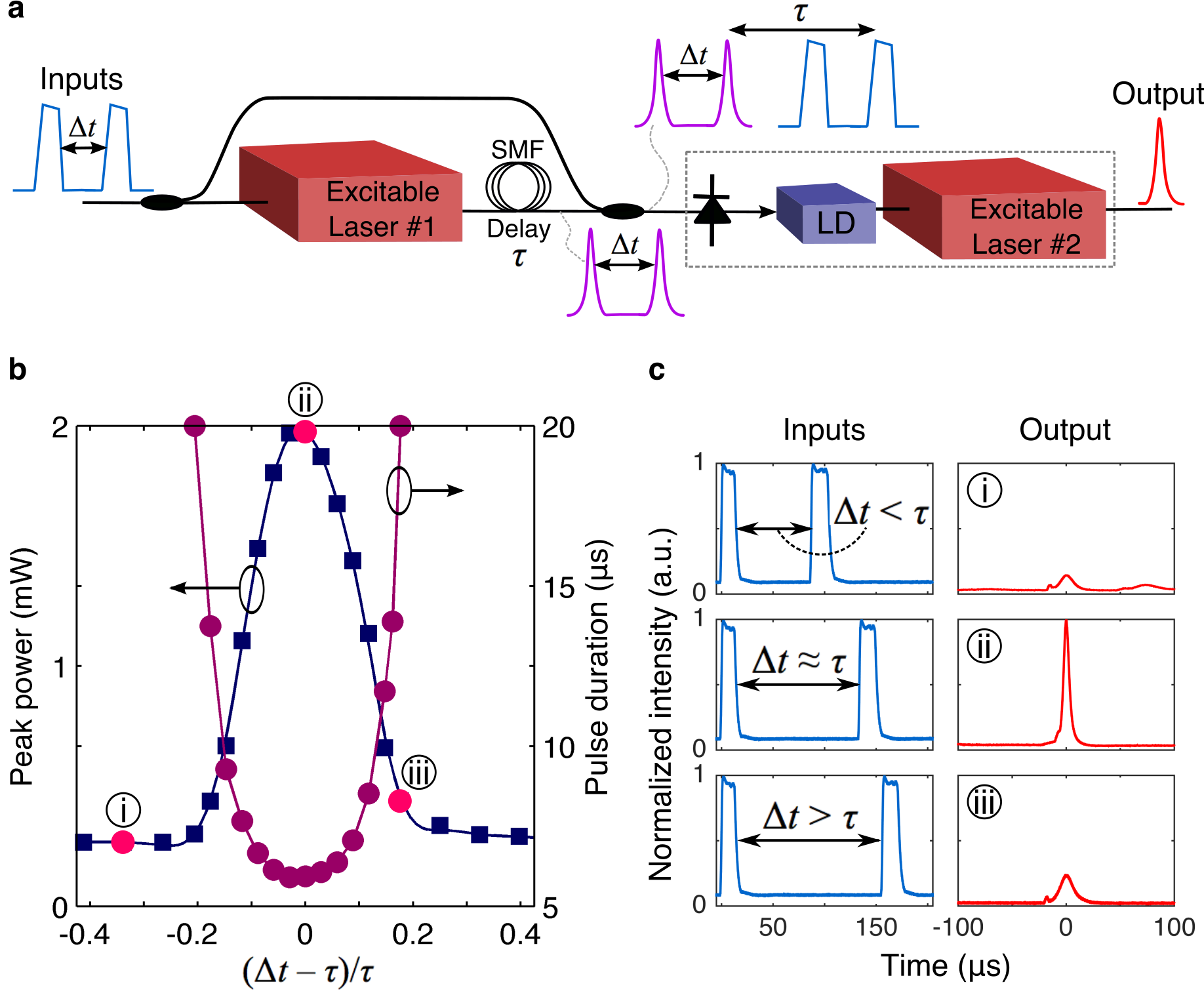}\caption{\label{fig:temporal_classifier}\textbf{Temporal pattern recognition.} \textbf{(a)} Simple circuit with two cascaded graphene excitable lasers. \textbf{(b)} Measured output pulse peak power and pulse duration as a function of the time interval between the two input pulses. \textbf{(c)} Measured input and output waveforms at specific instances: (i) $\Delta t-\tau=-45\mu s$, (ii) $\Delta t\approx\tau=135\mu s$, and (iii) $\Delta t-\tau=35\mu s$. The output pulse energy is largest when $\Delta t\approx\tau$ showing the system only reacts to a specific spatiotemporal input pattern.}
\end{figure*}

Since excitable systems are self-triggered, they exhibit important restorative properties. Different input perturbations often result in the same output, an important criteria for cascadability. Figures~\ref{fig:excitable_dynamics}e and f illustrate the response of the device as a result of a variety of input pulses. The excitable system responds in a stereotyped and repeatable way; all emitted pulses having identical pulse profiles. Outputs trigger asynchronously from input pulses, preserving analog timing information.

Figure~\ref{fig:excitability_properties} show some key behaviors associated with excitability. Fig.~\ref{fig:excitability_properties}a provides information about the refractory period for both the fiber laser and integrated device, which sets an upper bound on the pulse rate for a given unit. Similarly, Fig.~\ref{fig:excitability_properties}b shows the output pulse width as a function of an input pulse for both the integrated and fiber lasers. The fiber experiment is corroborated with matching simulation results (see Methods). Although the pulse profile stays the same, its amplitude may change depending on the value of the perturbation. The integrated device exhibits the same behavior on the much faster time scale, recovering in $\sim1$ nanosecond with pulse widths in $\mu$s, a factor of about $\sim 10^6$ faster than the fiber prototype in both respects.

Temporal pulse correlation is an important processing function that emerges from excitability. A integrating excitable system is able to sum together multiple inputs if they are close enough to one another in time. This allows for the detection of pulse clusters, or potentially, coincidence detection of pulses across channels through the use of incoherent optical summing.\citesup{Tait:jlt:2014} Coincidence detection underlies a number of processing tasks, including associative memory,\citesup{Markram:1997} and a form of temporal learning called spike timing-dependent plasticity (STDP).\citesup{Froemke:2002aa,Abbott:aa} Temporal pulse correlation in the fiber laser experiment and simulation, and integrated laser simulation are shown in Fig.~\ref{fig:excitability_properties}c. Reducing the time interval between input pulses (i.e. simultaneous arrival) results in an output pulse. Although the fiber laser can function at kHz speeds, the internal dynamics of the integrated device allow it to function much faster, putting it in the GHz regime.

{\flushleft{\bf Temporal pattern recognition.}} We demonstrate a simple pattern recognition circuit using several interconnected graphene fiber lasers. Pattern recognition of spatiotemporal phenomena is critical in the real-time processing of analog data. In the context of biological neural systems, networks of spiking neurons convert analog data into a spikes and recognize spatiotemporal bit patterns.\citesup{mohemmed2012span} Spatiotemporal patterns play an important role in both visual\citesup{Pillow:2008} and audio\citesup{Theunissen:2001} functionalities, and underly the formation of polychronized groups in the context of learning.\citesup{Izhikevich:2006}

As shown in Fig.~\ref{fig:temporal_classifier}a, we construct a simple two-unit pattern recognition circuit by cascading two excitable graphene (sic.) lasers with a delay $\tau$ between them. In our case the objective is to distinguish (i.e. recognize) a specific input pattern: a pair of pulses separated by a time interval $\Delta t\approx\tau$, equal to the delay between the excitable lasers. Coincidence detection provides the discriminatory power for classification.

A pulse doublet travels to both lasers, created using a modulator and arbitrary waveform generator. The output from the first laser travels to the second through a long single mode fiber ($\sim$km), which acts as a delay element. The second laser is biased with a larger threshold such that it will not fire unless two excitatory pulses---the original input and output from the first laser---arrive at the same time ($\Delta t\approx\tau=135\mu$s) (see Fig.~\ref{fig:temporal_classifier}b). Synchronous arrival of these two spikes causes the release of a pulse. Experimental time traces for the inputs and outputs are shown in Fig.~\ref{fig:temporal_classifier}c. A well-formed output pulse appears only for the desired two-pulse pattern.

One can reduce the occurrence of pulses that are at a non-normalized amplitude through a sharper threshold function (i.e. Fig.~\ref{fig:temporal_classifier}b). This ratio can be optimized for application-specific purposes. The first laser acts as a nonlinear stage to simply regenerate input pulses because it is biased close to the excitable threshold. The second laser, on the other hand, requires two coincident pulses to reach its threshold. It therefore plays the role of a pattern classifier. Between the laser stages, a photodetector (PD), rather than direct optical input, modulates the laser driver (allowing wavelength conversion from 1560 to 1480 nm). This PD-driven architecture (see outlined dashed box in Fig.~\ref{fig:temporal_classifier}a) has been explored in an integrated context\citesup{Nahmias:2014}, as a potential route to scalable on-chip networking.\citesup{Tait:jlt:2014} The dynamics introduced by the PD are analogous to synaptic dynamics governing the concentration of neurotransmitters in between signaling biological neurons.\citesup{Nahmias:2015} This simple circuit demonstrates several important features necessary for robust optical processing: well isolated input/output ports allow for the construction of feedforward networks, and the spatio-temporal recognition of spikes allows the system to classify patterns. More complex recognition and decoding would be possible as the system is scaled.

{\flushleft{\bf Stable recurrent circuit.}} We also demonstrate a self-recurrent graphene laser that can sustain a pulse traveling around the loop ad infinitum, providing a proof-of-principle demonstration of cascadability and pulse regeneration. Recurrently-connected dynamical networks which evolve toward a stable pattern over time (i.e. attractor networks) can exhibit hysteresis, and play a critical role in memory formation and recall.\citesup{Durstewitz:2000} Equivalently, since a single unit with a self-referent connection can be mapped to an infinite chain of lasers, this system can be viewed as a demonstration of stability in arbitrarily many layered feed-forward networks.

\begin{figure}[t]
\centering\includegraphics[width=1\columnwidth]{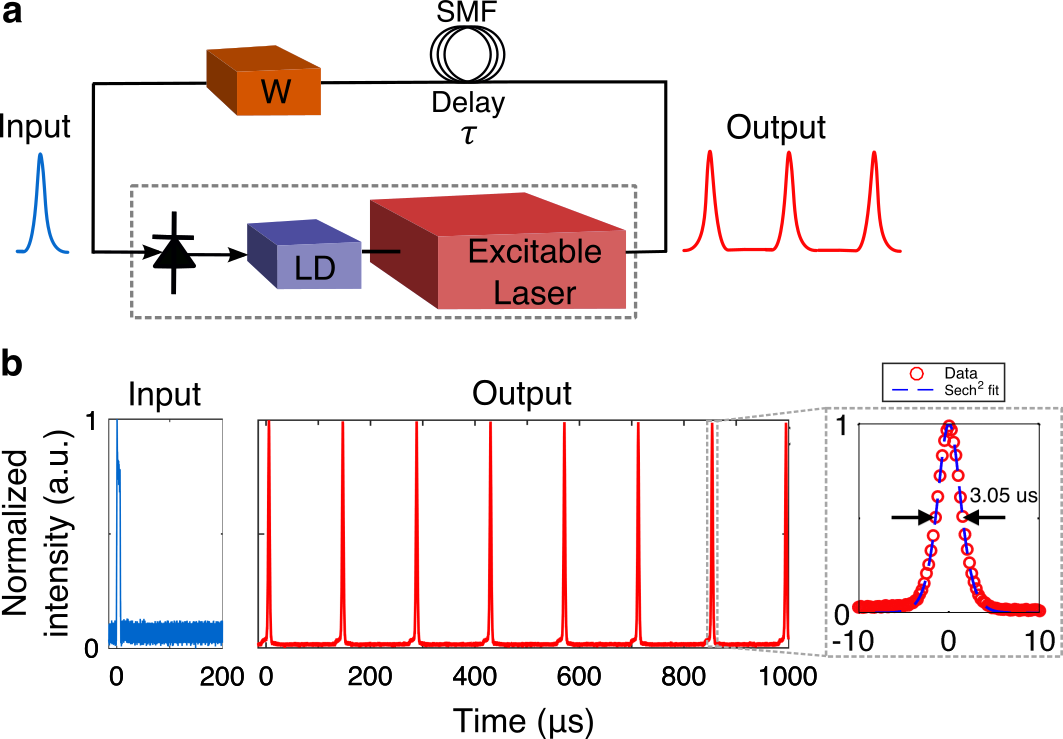}
\caption{\label{fig:recurrent_circuit}\textbf{Self-recurrent bistable circuit with the graphene excitable laser.} \textbf{(a)} Setup to test the self-referent connection. \textbf{(b)} Input and output waveforms. The first output pulse is fed back to the input after being delayed by $\sim$100 $\mu$s, which initiates another excitatory pulse at the output. This recursive process results in a train of output pulses \emph{ad eternum} at fixed intervals. Inset shows an output pulse profile and sech\textsuperscript{2} fitting curve.}
\end{figure}

Figure~\ref{fig:recurrent_circuit}a illustrates an excitable graphene laser with a self-referent connection. The output is fed back to the input via single-mode fiber which acts as a delay element (100 $\mu$s) with an electronic weight $W$ controls the modulation depth of the PD providing an all-or-none response depending on whether it is above or below a given threshold. Figure~\ref{fig:recurrent_circuit}b depicts the system's ability to demonstrate bistability when feedback is present. It is capable of settling to an attractor in which a single pulse travels around the loop indefinitely. This circuit represents a test of the network's ability to handle recursive feedback, and the stability of the pulse is a sign that the system is cascadable.

\section*{Discussion}\label{sec:discussion}
We have demonstrated that the complex dynamics of graphene excitable lasers can form a fundamental building block for spike information processing. In addition to single-laser excitability, we showed two instances of key spike processing circuits: temporal pattern recognition and stable recurrence. A photonic coincidence detection circuit forms the building block of the spatiotemporal pattern recognition circuit we have also demonstrated by cascading two excitable lasers as computational primitives. This simple demonstration of temporal logic implies that spiking neural networks of such excitable lasers are capable of categorization and decision making. Combined with learning algorithms such as STDP, networks could potentially perform more complex tasks such as spike-pattern cluster analysis.\citesup{Izhikevich:2006} A bistable recurrent spiking circuit enabled by the graphene excitable laser shows that processing networks of excitable lasers are capable of indefinite cascadability and information retention, a pre-requisite for more complex types of temporal attractors in recurrent networks. In networks of more lasers, spiking attractors can be more numerous, complex, and even competitive in order to achieve different information processing goals.

Ongoing research on graphene microfabrication could make it a standard technology accessible in integrated platforms. We proposed an integrated graphene-embedded cavity design and adapted the fiber model of excitability to a semiconductor device model. Our results show that an integrated device could maintain the essential behaviors required for spike information processing while reaping significant energy and speed improvements, potentially opening up applications for biologically inspired adaptive algorithms in presently inaccessible regimes of computing.\citesup{Nahmias:2013}

\section*{Methods}\label{sec:methods}
{\flushleft\bf{Fiber laser simulation.}} To simulate the fiber laser, we constructed rate equations based on the carrier dynamics in an EDF amplifier, roundtrip intensity, and loss. The dynamics of an EDF can be described using the following equations for fractional excited state population $n_2$, fractional ground state population $n_1$, and $k$ optical beams of intensities $I_k$\citesup{Giles:1991}:

\begin{align}
\frac{\partial n_2}{\partial t} =& \sum_k \frac{\sigma_{ak} I_k}{\hbar\omega_k} n_1(z,t) - \sum_k \frac{\sigma_{ek} I_k}{\hbar \omega_k} n_2(z,t) -\frac{n_2(t)}{\tau_{n}}
\end{align}
Each term represents the transition rate for each photon, where where $\hbar$ is Planck's constant and $\omega_k$ is the frequency of mode $k$. $\sigma_{ak}$ and $\sigma_{ek}$ represent the absorption and emission cross sections of each mode $k$, respectively, and the fractional populations satisfy $n_1 = 1 - n_2$. Our interest is in the modes at pump wavelengths 980 nm and 1480 and lasing modes which hover around 1520--1530 nm. We define \emph{pump intensity} $I_{p}$ at 980 nm, and \emph{input signal intensity} $I_{s}$ at 1480 nm, and the \emph{round trip intensity} $I_r$ at 1550 nm. Although the fiber laser is largely multi-mode, the modes are closely spaced to one another and possess similar cross sections. We can therefore approximate these modes with a single roundtrip intensity $I_ = \sum_k I_{k}$ equal to the sum of lasing modes, and define effective cross sections $\sigma_{er}$ and $\sigma_{ar}$. We also use a lumped approximation and represent the carrier density as a single variable.\citesup{Sun:1997} We average over the fiber length $z$ to arrive at the following differential equation for average carrier density $\overline{n_2} = \int n(z) dz$:
\begin{align}
\label{eq:avg}
\frac{d \overline{n_2}}{d t} =& \frac{\sigma_{ap} \overline{I_p}}{\hbar\omega_p} \overline{n_1}(t) + [\sigma_{as} \overline{n_1}(t) - \sigma_{es} \overline{n_2}(t)] \frac{\overline{I_s}}{\hbar\omega_s} \nonumber \\
&+ [\sigma_{ar} \overline{n_1}(t) - \sigma_{er} \overline{n_2}(t)] \frac{\overline{I_r}}{\hbar\omega_r} -\frac{\overline{n_2}(t)}{\tau_{n}}
\end{align}

Powers $P_k$ injected into the erbium-doped section are related to the average intensity within this section via $\overline{I_k} = \eta_k [(e^{g_k(t)}-1)/A_{eff} g_k(t)] P_k (t)$, where $g_k(t) \approx \Gamma_k n_t [\sigma_{ek} \overline{n_2}(t) - \sigma_{ak} \overline{n_1}(t)] L_{Er}$ is the gain experienced by the mode over the length of the erbium fiber, $A_{eff}$ is the effective cross sectional area of the fiber, $\eta_k$ is the injection efficiency, $\Gamma_k$ is the confinement factor, $n_t$ is the erbium ion density, and $L_{Er}$ is the length of the erbium section.\citesup{Sun:1997}

We can define roundtrip equations for both round-trip loss $q(t)$ and round-trip power averaged over the fiber length $P_r(t)$:
\begin{align}
\label{eq:q}
\frac{d q}{dt} &= - \frac{q(t) - q_0}{\tau_q} - \frac{q(t) P_r(t)}{E_{sat}} \\
\label{eq:Pr}
T_R \frac{dP_r}{dt} &= [g_r(t) - q(t)  - l]P_r(t) + \rho_{sp}
\end{align}
where $q_0$ represents the small-signal absorption of the SA, $\tau_q$ the absorber relaxation time, $E_{sat}$ the saturation energy, $T_R$ the round-trip cavity time, $l$ the round-trip loss, $\rho_{sp}$ a small spontaneous noise term, and $g_r(t) = \Gamma_r n_t [\sigma_{er} n_2(t) - \sigma_{ar} n_1(t)] L_{Er}$ the erbium fiber round-trip gain. Equations (\ref{eq:avg}--\ref{eq:Pr}) represented the model used for the simulation. Parameters are shown in Table \ref{tab:fiber_param}. These equations were stepped iterately using Runge-Kutta methods to generate time traces and measure various properties.

\begin{table}[!t]
\renewcommand{\arraystretch}{1.2}
\caption{Fiber Laser Param. (from data and \cite{Bao:2009})}
\label{tab:fiber_param}
\centering
\begin{tabular}{lll}
\hline\hline
{\bfseries Param.} & {\bfseries Description} & {\bfseries Value}\\
\hline
$I_p$ & pump intensity & $4.39 \times 10^8$ W/$\mathrm{m}^2$ \\
$\eta_k$ & pump coupling efficiency & .21\\
$\nu_s, \nu_p$ & input sig. and pump freq. & 194, 306 THz \\
$\tau_n$ & erbium lifetime & 9 ms\\
$\sigma_{ap}$ & abs. cross section (pump) & $2.87 \times 10^{-25}$ $\mathrm{m}^2$\\
$\sigma_{as}, \sigma_{es}$ & abs., em. cross sec. (sig.) & 3.01, .948 $\times 10^{-25}$ $\mathrm{m}^2$\\
$\sigma_{ar}, \sigma_{er}$ & abs., em. cross sec. (rt.) & 3.21, 4.54 $\times 10^{-25}$ $\mathrm{m}^2$\\
$A_{eff(r)}$ & eff. cross sec. area (rt.) & $2.77 \times 10^{-11}$ $\mathrm{m}^2$\\
$\Gamma_{k(p)}, \Gamma_{k(s)}$ & conf. factors (pump, sig.)  & .849, .638\\
$n_t$ & erbium ion density & $5.8 \times 10^{25}$ $\mathrm{m}^3$\\
$L_{Er}$ & erbium fiber length & 75 cm\\
$q_0$ & small-sig. SA absorption & .5\\
$\tau_q$ & SA lifetime & 2 ps\\
$E_{sat}$ & SA saturation energy & 10 pJ\\
$T_R$ & cavity roundtrip time & 90 ns\\
$l$ & cavity intrinsic loss & 1.1\\
$\rho_{sp}$ & spont. noise term & 5 W/s\\

\hline\hline
\end{tabular}
\end{table}

It is possible to recover the simplified, undimensionalized model that underlies the observed behaviors by noting that $n_2 (t)$ does not change significantly over time (i.e. $n_2(t) = n_K + \delta n(t)$ where $\delta n(t) << n_K$) and substituting $g_r(t)$ into equation (\ref{eq:avg}). These approximations lead to a bilinear set of equations that are analogous to equations (\ref{eq:yamadaG}--\ref{eq:yamadaI}).

{\flushleft{\bf Integrated-device simulation.}} Design principles are chosen for compatibility with recent graphene deposition and patterning techniques.\citesup{Bonaccorso:2010aa,Kim:2011} The device epitaxial layer structure includes both quantum wells (QWs) and graphene coupled to a single optical mode, shown in Figure \ref{fig:integrated_laser}. Both the QWs and graphene provide complementary properties---whereas QWs provide high efficiency gain, the graphene provides strong, fast and wideband saturable absorption. The difficulty in coupling graphene directly to the optical mode could be resolved for instance by wafer bonding a III-V laser on top of deposited graphene, avoiding any interaction between graphene and electrical pumping, as shown. For improved dynamics, we consider two pristine layers of graphene, protected by a atomically flat layers of boron nitride (BN) to prevent each graphene sheet from interacting too strongly with surrounding materials. We computed the optical mode of this structure using an eigenmode expansion (EME) technique.

Using the confinement factor from above and other various parameters, we simulated the device using a lumped rate equation model. Beginning with the theory for graphene, the behavior is well approximated by a simple saturation model, given by\citesup{Bao:2009}:
\begin{equation}
	\alpha(\nu_{\alpha}) = \frac{\alpha_S}{1 +\nu_{\alpha}/\nu_{s}} + \alpha_{NS}
\end{equation}
where $\alpha(\nu_\alpha)$ represents the absorption coefficient (per unit length), $\nu_\alpha$ is the two-dimensional carrier density in graphene, $\nu_s$ the 2D saturation carrier density, $\alpha_{NS}$ the saturable absorption, and $\alpha_{NS}$ the non-saturable absorption. The resulting rate equations are given by:

\begin{subequations}
\label{eq:gen}
\begin{align}
\frac{dN_{ph}}{dt} &= v_g g(n_g) N_{ph} - v_g a(\nu_\alpha) N_{ph} + \frac{N_{ph}}{\tau_{ph}} + R_{sp} \\
\frac{dn_g}{dt} &=  \frac{I_g + \phi(t)}{e V_g} - v_g g(n_g) \frac{N_{ph}}{V_g} - \frac{n_g}{\tau_g} + \theta(t) \\
\frac{d\nu_\alpha}{dt} &= - \frac{\nu_\alpha}{\tau_a} + v_g a(\nu_\alpha) \frac{N_{ph}}{A_\alpha}
\end{align}
\end{subequations}
$N_{ph}$ represents the number of photons in the cavity and $n_g$ the carrier density QW gain region. (Note: the variable $\nu_\alpha$ represents the \emph{surface} carrier density within graphene, chosen for convenience as graphene is two-dimensional.) $g(n_g)$ and $a(n_\alpha)$ describe the gain and absorption per unit length, $v_g$ the group velocity, $\tau$ the lifetimes, $I_g$ current pumped into the gain region, $R_{sp}$ a small spontaneous noise term, $V_g$ the volume of the gain region, $A_\alpha$ the area of the graphene sheet, and $\phi(t)$ an input current modulation term. The input power $P_{g}$ that is driving the laser can be computed by $P_{g} = I_g \times v_L$ where $v_L$ is the voltage applied across the gain section of the laser.
Gain and loss are assumed to take the forms:
\begin{align*}
g(n_g) &= \Gamma_g g_0 \log(n_g/n_{tr}) \\
a(\nu_{\alpha}) &= \frac{\alpha_S}{1 +\nu_{\alpha}/\nu_{s}}
\end{align*}
where $\Gamma_g$ is the gain confinement factor, $n_{tr}$ is the transparency density ($\mathrm{cm}^{-3}$) and $\nu_{s}$ is graphene transparency density in two dimensions ($\mathrm{cm}^{-2}$). Non-saturable absorption $\alpha_{NS}$ is not included as it manifests as cavity losses, becoming absorbed into the photon lifetime $\tau_{ph}$. Parameters are shown in Table \ref{tab:int_param}. We simulated the rate equation model using Runge-Kutta methods.

\begin{table}[!t]
\renewcommand{\arraystretch}{1.2}
\caption{Hybrid Integrated Laser Param. (from simulations, and \cite{Fang:2007,Zhang:2014low})}
\label{tab:int_param}
\centering
\begin{tabular}{lll}
\hline\hline
{\bfseries Param.} & {\bfseries Description} & {\bfseries Value}\\
\hline
$\lambda$ & lasing wavelength & 1550~nm\\
$v_g$ & group velocity & $c$/3.5 \\
$V_g$ & gain region volume & 2.55$\times$10$^{-\textnormal{12}}$ cm$^{\textnormal{3}}$\\
$A_\alpha$ & graphene sheet area & 1.5$\times$10$^{-\textnormal{6}}$ cm$^{\textnormal{2}}$\\
$\Gamma_g$ & gain region confinement factor & 0.034\\
$\tau_g$ & gain region carrier lifetime & 1.1~ns\\
$\tau_\alpha$ & graphene carrier lifetime & 405 fs\\
$\tau_p$ & photon lifetime & 2.4 ps\\
$g_0$ & log gain coefficient & 972 cm$^{-\textnormal{1}}$\\
$\alpha_s$ & waveguide saturable absorption & 150 cm$^{-\textnormal{1}}$\\
$n_{tr}$ & 3D gain region transparency density & 1.75$\times$10$^{\textnormal{18}}$ cm$^{-\textnormal{3}}$\\
$\nu_{s}$ & 2D graphene transparency density & 1.06$\times$10$^{\textnormal{13}}$ cm$^{-\textnormal{2}}$\\
$R_{sp}$ & spontaneous noise term & $1 \times 10^{10}$ $\mathrm{s}^{-1}$\\
$v_L$ & applied voltage (gain section) & 1.1 V\\
\hline\hline
\end{tabular}
\end{table}

One can recover the undimensionalized equations with several approximations and variable substitutions. Making a linear approximations to both the gain and absorption, i.e. $g(n_g) \approx g_0 [n_g - n_{tr}]/n_{tr}$ and $a(\nu_\alpha) \approx \alpha^*_S [1 - \nu_\alpha/\nu_s]$, leads to an equation analogous to the simplified gain-absorber model described by equations (\ref{eq:yamadaG}-- \ref{eq:yamadaI}).

{\flushleft\bf{Excitable fiber ring laser cavity.}} The EDF employed in the laser cavity is a gain fiber (LIEKKI Er80-4/125), with peak core absorption coefficients of 60, 50 and 110 dBm$^{-1}$ at 980, 1480, and 1530 nm, respectively. It has a large area core with a mode field diameter of 6.5 $\mu$m at 1550 nm and a core numerical aperture of 0.2. The high erbium ion doping concentration reduces the required fiber length significantly while providing strong gain and reducing nonlinear effects (four-wave mixing, stimulated Raman scattering, stimulated Brillouin scattering). The length of the EDF (75~cm) is chosen to ensure population inversion with the desired pump power so that the EDF does not play any role as a SA to realize excitability. All fibers used in the cavity are polarization-independent. A polarization controller consisting of three spools of SMF-28 fibers acting as retarders is used to maintain a given polarization state after each round trip improving the output pulse stability.\citesup{Popa:2011}

{\flushleft{\bf Graphene sample preparation.}} Graphene samples are prepared by chemical reduction of graphene oxide (GO) with hydroxylamine hydrochloride (NH\textsubscript{2}OH\textsubscript{3}$\cdot$HCl) with a slightly modified recipe.\citesup{Zhou:2011} 25 ml of 0.5 mg/ml GO (Graphene supermarket \#SKU-GO-W-175) is diluted with 25 ml of deionized (DI) water, 200 $\mu$L of 28 wt.\% ammonium hydroxide (Sigma-Aldrich \#338818), and 25 mg of NH\textsubscript{2}OH\textsubscript{3}$\cdot$HCl (Sigma-Aldrich \#431362) in a 100 ml round-bottom flask and stirred. The mixture is transferred to a water bath and heated at $\sim$90\textsuperscript{o}C with stirring at 350 rpm for 90 mins. The color of the mixture changes from yellowish brown to homogeneous black and precipitating from solution indicating reduction has taken place. The reduced GO (rGO) is filtered and washed three times with DI water. The rGO is then suspended in 50 ml of DI water with 50 mg of sodium deoxycholate (Sigma-Aldrich \#30970), and stirred until the salt dissolves. The mixture is sonicated for 30 mins using ultrasound-assisted functionalization resulting in a stable rGO suspension. A micropipette is used to transfer 5 $\mu$L of the as prepared liquid to an angle-polished fiber connector (FC/APC). The deposited sample is dried with a heat gun operating at around 120\textsuperscript{o}C for $\sim$3 to 5 mins.

\section*{Acknowledgements}
B.J.S acknowledges the support of the Banting Postdoctoral Fellowship administered by the Government of Canada through the Natural Sciences and Engineering Research Council of Canada. M.A.N and A.N.T acknowledge the support of the the National Science Foundation Graduate Research Fellowship.

\section*{Author Contributions}
B.J.S., M.A.N., and A.N.T. contributed equally to this work. B.J.S. and A.N.T. designed and performed the experiments. M.A.N. and A.W.R. developed the theoretical model and performed the simulations. B.J.S. fabricated the samples. B.W. helped with the experiments. P.R.P. supervised the project. All authors contributed to the discussion of the results and implications, and to the preparation of the manuscript.


\bibliographystyle{naturemag}

\clearpage

\setcounter{figure}{0}
\renewcommand{\figurename}{Figure S}
\section*{Supplementary Figures}

\begin{figure}[htbp]
\centering\includegraphics[width=1.0\columnwidth]{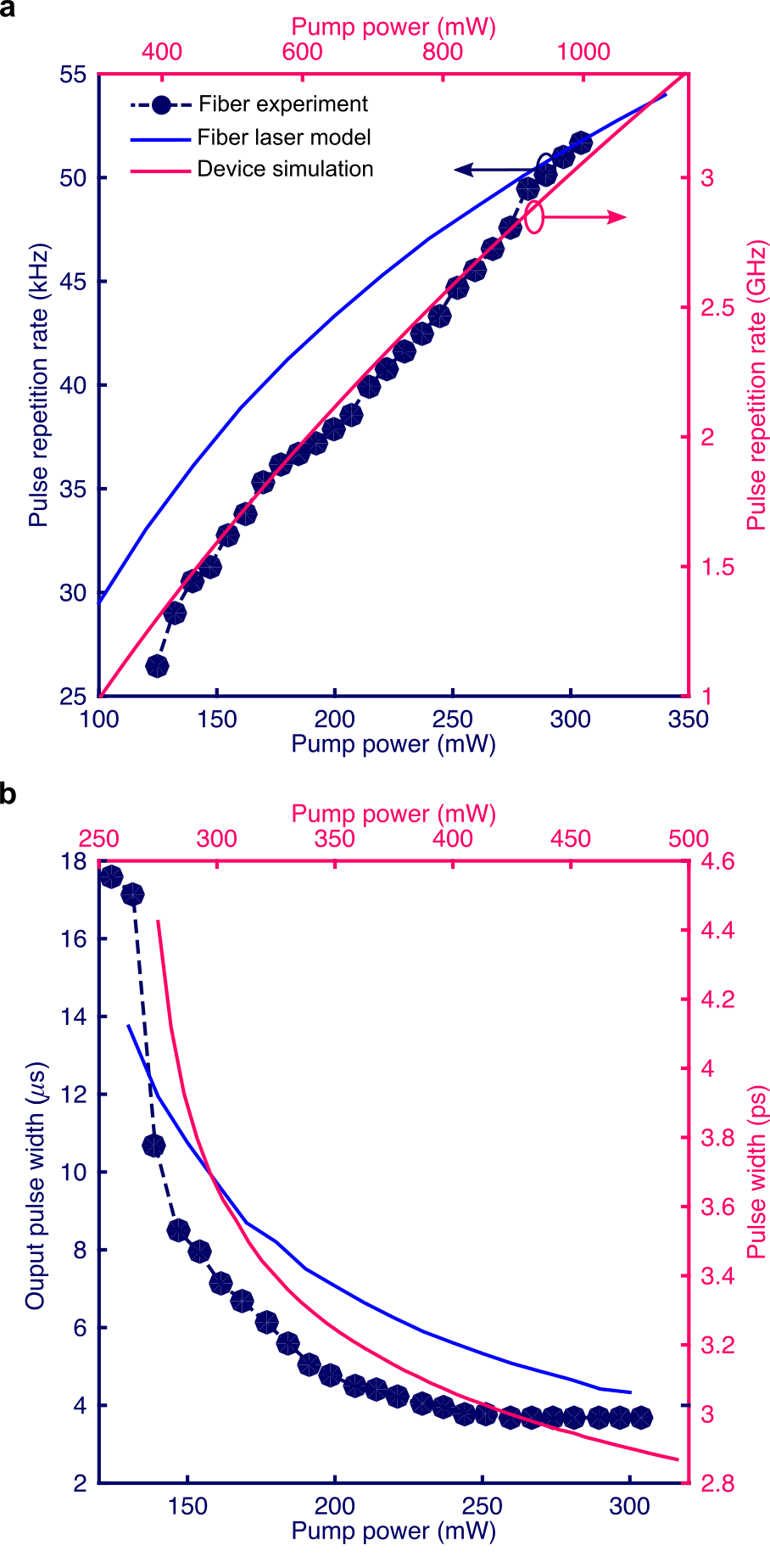}
\caption{\label{fig:q_switching} Typical characteristics of the passively Q-switched fiber and integrated lasers. (a) Output pulse repetition rate and (b) pulse width as a function of the pump current.}
\end{figure}

\end{document}